\documentclass[aps,prl,twocolumn,preprintnumbers,
superscriptaddress,floatfix]{revtex4}

\usepackage{amssymb,multirow,amsmath}
\usepackage{graphicx,color}

\begin{document}

\newcommand{\Tr}{\mbox{Tr\,}}
\newcommand{\beq}{\begin{equation}}
\newcommand{\eeq}{\end{equation}}
\newcommand{\bea}{\begin{eqnarray}}
\newcommand{\eea}{\end{eqnarray}}
\renewcommand{\Re}{\mbox{Re}\,}
\renewcommand{\Im}{\mbox{Im}\,}

\title{Domain Wall AdS/QCD}

\author{Nick Evans}
\affiliation{ STAG Research Centre \&  Physics and Astronomy, University of
Southampton, Southampton, SO17 1BJ, UK}

\author{Jack Mitchell}
\affiliation{ STAG Research Centre \&  Physics and Astronomy, University of
Southampton, Southampton, SO17 1BJ, UK}

\begin{abstract}
We construct a new holographic description of QCD using domain wall fermions. The construction consists of probe D7 branes in a D5 brane geometry describing quarks on a 4+1d defect in a 5+1d gauge theory.  We then compactify one dimension of the D5 to introduce confinement in the gauge degrees of freedom. In addition we allow a spatial dependent mass term for the D7 brane quarks to isolate chiral fermions on 3+1d domain walls. The D7 world volume fields, when restricted to the domain wall position, provide an AdS/QCD description. We compute the spectrum and compare to data. We include higher dimension operators to systematically improve the description.
\end{abstract}

\maketitle

Domain wall fermions \cite{Kaplan:1992bt} are a powerful technique for isolating massless, chiral fermions within a gauge theory. The technique is widely used in lattice QCD simulations to enforce chiral symmetry. Recently we investigated the technique in a holographic setting \cite{CruzRojas:2021pql} providing a holographic description of 2+1 dimensional domain wall fermions on a probe D7 brane in the AdS$_5$ space generated by ${\cal N}=4$ super Yang-Mills theory \cite{Maldacena:1997re}. In the limit where the higher dimension mass is very large the position of the domain wall (where the chiral fermions are massless) can be found exactly. Restricting the holographic fields to the locus of the domain wall gives a holographic description of the dynamics of those chiral fermions. 

Here we take this approach to provide a description of a 3+1 dimensional domain wall theory with $N_f$ chiral quarks on the defect - the basic construct is a 5+1 dimensional gauge theory (on a D5 brane) compactified in one dimension (introducing confinement), with quarks present on 4+1 dimensional defects (probe D7 branes). The domain wall structure is then used to place chiral fermion on 3+1 dimensional defects. When the 4+1d mass is large the position of the domain wall can be found and the holographic fields, when reduced to this locus, provide a description of the chiral fermions.   We present the construction of this Domain Wall AdS/QCD theory and compute the light meson spectrum it predicts. The UV of the theory, reflecting that the gauge dynamics is 5+1 dimensional, does not match to perturbative QCD so we impose a cut off at the 3 GeV scale and only work at lower scales in the holographic model. The predictions are comparable in quality to those of other AdS/QCD constructions \cite{Erlich:2005qh,Sakai:2004cn}. 

The holographic description should be matched at the 3 GeV upper cut off to QCD in the intermediate coupling regime and higher dimension operators would be expected to be present \cite{Evans:2006ea}. We include such operators using Witten's multi-trace prescription \cite{Witten:2001ua} (see \cite{Evans:2016yas,Clemens:2017udk} for previous examples of using HDOs in holographic descriptions of QCD). We fit the couplings of these operators to the meson data since we can not compute the non-perturbative QCD matching.  We show that the predictions of the model can be systematically improved in this way.  \vspace{-0.6cm}

\section{I The Brane Construction} \vspace{-0.3cm}

Our construction is built around the D5/probe D7 system with five coincident directions in the configuration (one of the systems discussed in \cite{Myers:2006qr}). 
 
      \begin{center}
 \begin{tabular}{c|c c c c c c c c c c}
      &0&1&2&3&4&5&6&7&8&9   \\   \hline
      D5&-&-&-&-&-&(-)&$\bullet$ &$\bullet$ &$\bullet$ &$\bullet$ \\
      D7&-&-&-&-&-&$\bullet$  &-&-&- &$\bullet$ 
 \end{tabular}
\end{center} \vspace{-2cm} 

\beq \label{one}\eeq \vspace{0.1cm}

The UV theory is therefore a supersymmetric 5+1d gauge theory with quark hypermultiplets restricted to a 4+1d domain wall. The gauge theory is strongly coupled in the UV but we will set up our QCD-like dynamics in the IR where the supergravity approximation holds. We will compactify one of the five spatial directions on the D5 brane, shown by the brackets in (\ref{one}). This breaks supersymmetry and introduces an IR confinement scale by making the geometry a cigar in the $x_5$ and radial direction. 

Note if the D7 brane were at $x_9=0$ describing a massless quark, then the D7 would wrap around the cigar and re-emerge as an anti-D7 brane anti-podal on the circle in $x_5$. This demonstrates that the theory needs an anti-D5 in order for the D7 fluxes to have a sensible solution on the $x_5$ circle. Here though we will, except on a domain wall, set the quark mass very large so that the D7 only live at large radius where they are widely separated on the $x_5$ circle. We will assume that there is then no interaction between the anti-podal  branes and concentrate on the dynamics on one brane. 

The final trick we will employ is to allow the quark mass, $M$, on the 4+1d defect to be $x_4$ dependent. We will assume it is positive and very large everywhere except in an interval of width $w$ where the sign flips. The boundaries of this region have $M=0$ and are the domain walls. One expects the localization of 3+1d chiral fermions, one on each domain wall. As in the previous examples we studied in \cite{CruzRojas:2021pql}, the domain walls approach and merge as one moves into the IR of the holographic description indicating the presence of chiral symmetry breaking. The $v_9$ field, which describes the condensate of the left and right handed fermions, can have solutions isolated on the domain wall. We show that the solutions display, consistently, chiral symmetry breaking solutions on the U-shaped embeddings of the domain wall configurations. \vspace{-0.5cm}

\subsection{D5 Geometry} 

The geometry generated by the D5 branes with a compact direction is known. One takes the thermal geometry for example found in \cite{Itzhaki:1998dd} and Wick rotates to interchange a spatial and the time direction as described in \cite{Horowitz:1998ha}. This leads to the near-horizon  metric ($U=r/\alpha'$, $K= \frac{(2\pi)^{3/2}}{g_{YM}\sqrt{N}}$)
   \beq 
   {ds^2 \over \alpha'} =  K U(- dt^2 + dx_{4}^2 + h dz^2)~~ +   \frac{1}{K U}\Big({1 \over h}dU^2 + U^2d\Omega_3^2\Big)  \eeq
   where 
   \beq
   h(U) = 1- {U_0^2 \over U^2}\eeq\beq
   e^\phi ={U \over K}, ~~~~~~ g_{YM}^2=(2\pi)^{3}g_s \alpha'   \eeq
   Note that in the 5+1d dual the gauge field is of energy dimension one so $1/g^2_{YM}$ has energy dimension two.  Here we see that $U$ has dimension one and the dilaton is dimensionless. 
   
   To find the circumference of the circle in $z$ we expand near the ``horizon'', $U=U_0+\delta U$, and find to leading order in the $U-z$ plane 
   \[ ds^2 = 2 K \delta U dz^2 + {1 \over 2 K \delta U} d \delta U^2 \]
   We then set $\alpha =  K z$ and $\delta U= {1\over 2} K \sigma^2$ and obtain
   \[ ds^2 =  \left( d \sigma^2 + \sigma^2 d \alpha^2 \right) \]
   which is a plane. To have no deficit angle $0 < \alpha < 2 \pi$ so $0<z< 2 \pi/K$.
   
   Before we can embed the D7 brane we need to write the metric so that the directions transverse to the D5 are a flat plane (as in the cases explored in \cite{Babington:2003vm}). The relevant pieces of the metric are 
   \beq
   ds^2 =  \frac{U }{K} \Big({1 \over U^2 h(U)}dU^2 + d\Omega_3^2\Big) \eeq
   We change coordinates so 
   \beq {dv^2 \over v^2} = {dU^2 \over U^2 h(u)} \label{veqn} \eeq
   so that 
   \beq
   ds^2 =  \frac{1}{K} {U(v)\over v^2} \Big(dv^2 + v^2 d\Omega_3^2\Big) \eeq
   which we can then write as 
    \beq
   ds^2 =  {1 \over K}{U(v)\over v^2} \Big(d\rho^2 + \rho^2 d\Omega_2^2 + dv_9^2\Big) \eeq
   
   Solving (\ref{veqn}) gives 
   \beq v^2 =  {1 + {u \over \sqrt{u^2-u_0^2}} \over  {u \over \sqrt{u^2-u_0^2}}-1} ~~~~ {\rm or} ~~~~
       {U \over U_0}={1+v^2 \over 2v}
   \eeq
   Note that  $v \rightarrow 1$ as $U/U_0 \rightarrow 1$ and at large $U$ we find
   $v^2 = 4 U^2/U_0^2$.

   Finally, the metric can be written
   \beq  ds^2 =  G_x ( dx_{0-3}^2 + h dx_5^2) + {G_v } (d\rho^2 + \rho^2 d \Omega_2^2 + dv_9^2) \eeq
with
\beq G_x= K U_0 {v^2+1 \over 2 v}, \hspace{1cm}  h(v) = 1 - \left( {2 v \over v^2 +1} \right)^2
\eeq
\beq G_v =  {U_0\over K}{1+ v^2\over 2v^3} \hspace{1cm} e^{-\phi} = {K \over U_0} {2v \over 1+v^2} \eeq
   
   It is worth noting here that the holographic directions in this set of coordinates do not carry the field theory energy dimensions. $G_x$ does and this is a useful check of equations below.

 \subsection{D7 Probe Action}
 
 Next we include 4+1d quark hypermultiplets on defects in the 5+1d glue theory by the inclusion of a probe D7 brane \cite{Karch:2002sh} in the configuration of (\ref{one}). The DBI action takes the form
   \beq S_{D7} = - T_7 \int d^8 \xi \sqrt {{\rm det} ( P[G_{ab}] + 2 \pi \alpha'F_{ab}) }\eeq
   where $\xi$ are the world volume coordinates and $P$ denotes the pullback. We find, setting the worldvolume vector to zero for now, 
\beq \label{d7act1} \begin{array}{r} S_{D7} = - {\cal N} \int d^5x~ d\rho ~ \rho^2 e^{-\phi} G_x^{5/2} G_v^{3/2} \hspace{0.5cm} \\ \\ \times \sqrt{1  + (\partial_\rho v_9)^2 + {G_v \over G_x} (\partial_{x_{0-4}} v_9)^2}
\end{array}\eeq
where ${\cal N}=T_7\  \int d \Omega_2$. The factor
\beq \label{factor}
e^{-\phi}G_x^{5/2} G_v^{3/2} = {K^2 U_0^3\over 8}\big(1 + {1 \over v^2}\big)^3
\eeq
and blows up as $v \rightarrow 0$ which encourages the D7 to bend away from $v=0$ by switching on $v_9$ and generating chiral symmetry breaking. 
The equation for the D7 embedding that encodes this is
\beq \begin{array}{l}\partial_\rho \left[ {\rho^2 e^{-\phi} G_x^{5/2} G_v^{3/2} \over \sqrt{1  + (\partial_\rho v_9)^2}} \partial_\rho v_9 \right] \\ \\  - 2 \rho^2 \sqrt{1  + (\partial_\rho v_9)^2} \left({d \over dv^2} e^{-\phi}G_x^{5/2} G_v^{3/2} \right) v_9 = 0 \end{array} \eeq
The UV solution is $v_9\simeq M + C/\rho ~(\simeq U/2U_0 )$ and so the mass is proportional to $M U_0$ and the condensate (of dimension four in 4+1d) to $C K^2 U_0^2$ (note that the condensate is a derivative with respect to the mass on the action so naturally picks up the $K^2$ factor from (\ref{factor})). We will avoid this chiral symmetry breaking (and any interaction with any anti-podal anti-D7) by taking configurations where $M \rightarrow \infty$ except on domain walls.

\subsection{Domain Walls}

Our final ingredient is to introduce a quark mass that has spatial dependence in the $x_4$ direction. We take the UV  mass to be $M$ except, on the boundary,
\beq \label{bigmass} v_9 = -M   \hspace{1cm}  -w/2  < x_4 <  w/2 \eeq
We expect 3+1d chiral fermions to be isolated at the two discontinuities where $M=0$. We will now work in the infinite $M$ limit \cite{CruzRojas:2021pql} so that any issues with the 4+1d quarks are pushed to the far UV and so that the $x_4$ derivative of $v_9$ becomes a delta function. One must be careful to include appropriate Jacobian factors in the form of the delta function (these are those that effectively reduce 
the D7 action to that of a 6 brane). We have, with $M$ vanishing on the contour $x_4(\rho)$
\beq \partial_\rho v_9 = \left. {1 \over G_v^{1/2} (\partial_4 \rho)}\right|_{\rm locus} \delta(x_4-x_4(\rho)) \label{delta1}\eeq

We now insert this factor into the D7 action (\ref{d7act1}) assuming that $v_9 =0$ (formally $v_9 \ll M$) giving
 \beq S_{\rm locus} = - {\cal N} \int d^4x ~d\rho~ \rho^2 e^{-\phi} G_x^{2} {G_v^{3/2} } \sqrt{1+{G_x \over G_v} (\partial_{\rho} x_4)^2} \label{eqlocus1}\eeq
 
   which is an action that determines the locus on which $M=0$ in the $\rho-x_4$ plane.   (\ref{eqlocus1}) has a conserved quantity which we denote ${\cal C}$ and we find
  \beq \partial_\rho x_4 = {G_v^{1/2} \over G_x^{1/2} \sqrt{e^{-2\phi}
\rho^4 G_x^5G_v^2 {\cal C}^2-1}}\eeq

 \begin{center}
    \includegraphics[width=8cm]{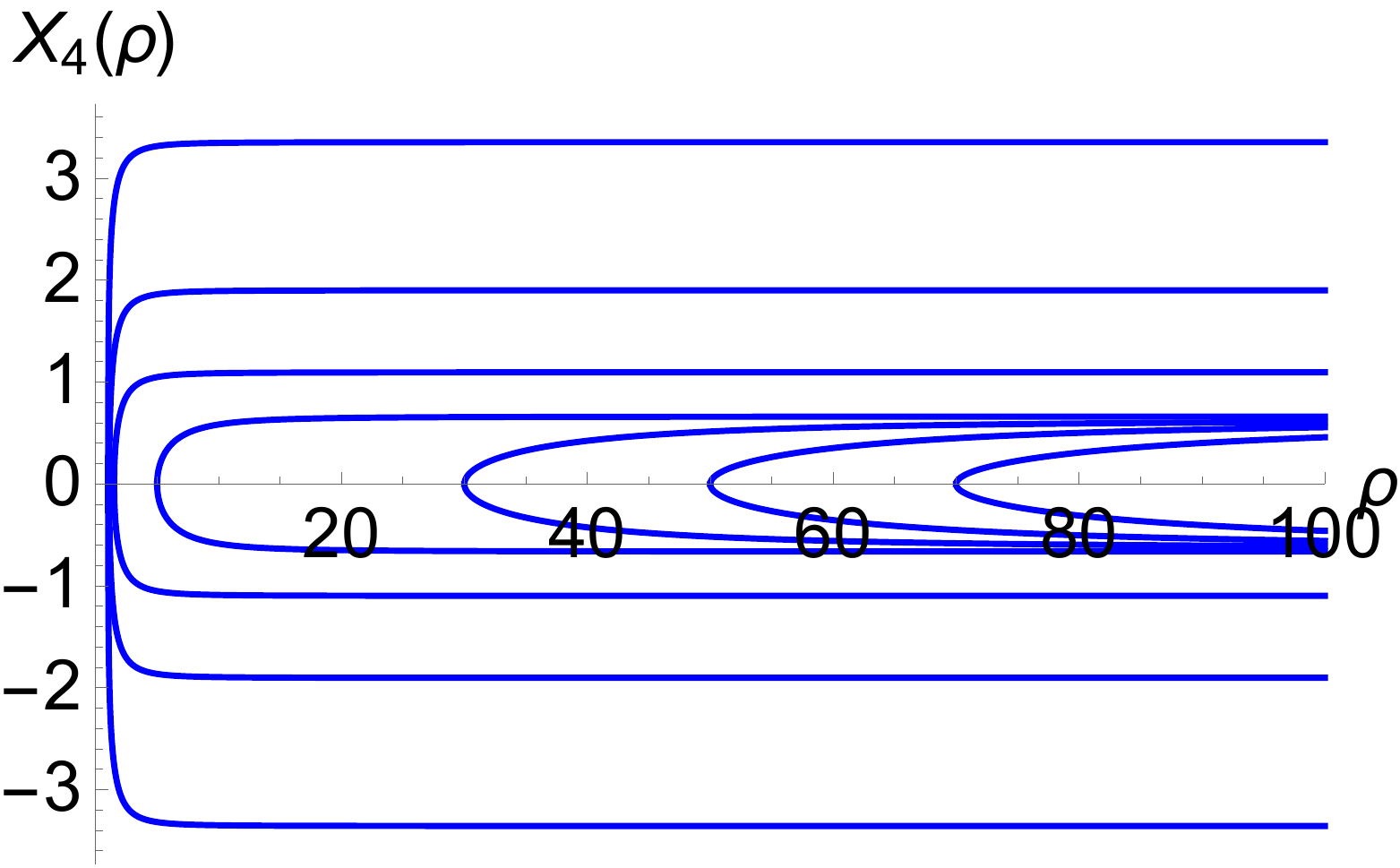} 
 \vspace{-0.4cm}
 
\noindent{{\textit{Figure 1: The loci  of the domain walls in the $\rho-x_4$ plane for different choices of ${\cal C}/\rho_{\rm min}$. Here we set $KU_0=1$ for numerics.}}}
\end{center} \vspace{-0.2cm}

Note the large $\rho$ limit of this is $4 \sqrt{2} /({\cal C} K^{7/2} U_0^{5/2} \rho^{7/2})$ and ${\cal C}$ has energy dimension -5. 

   The solutions are U-shaped in the $\rho-x_4$ plane with the minimum $\rho$ value given when the denominator vanishes. We display these solutions in Figure 1.

\section{II The Domain Wall Theory}

We now wish to describe holographically the 3+1d chiral fermions living on the domain walls and their interactions - this is the Domain Wall AdS/QCD theory. One wants solutions of the D7 brane world volume fields that are of the form of a delta function on the loci found above and shown in Figure 1. To find such solutions we, by hand, dimensionally reduce the D7 brane action in (\ref{d7act1}) onto the loci by imposing a delta function of the form in (\ref{delta1}).

\subsection{The Quark Mass and Condensate} 

As a first example let's find the vacuum configuration describing the quark condensate by considering just the field $v_9$. We obtain the action
\beq \begin{array}{r}S_{D7} = - {\cal N} \int d^4x ~d\rho ~ \rho^2 e^{-\phi} G_x^{5/2} {G_v^{3/2} \over G_v^{1/2}(\rho)}(\partial_\rho x_4) \\ \\\times \sqrt{1  + {\cal F}(\partial_\rho v_9)^2 +{G_v \over G_x} (\partial_{x_{0-3}} v_9)^2} \end{array} \label{mess}\eeq
where 
\beq {\cal F} = 1 + {G_v \over G_x (\partial_\rho x_4)^2}   \eeq

It's worth noting that in the large $\rho$ limit for the pieces

  \begin{center}
    \includegraphics[width=8cm]{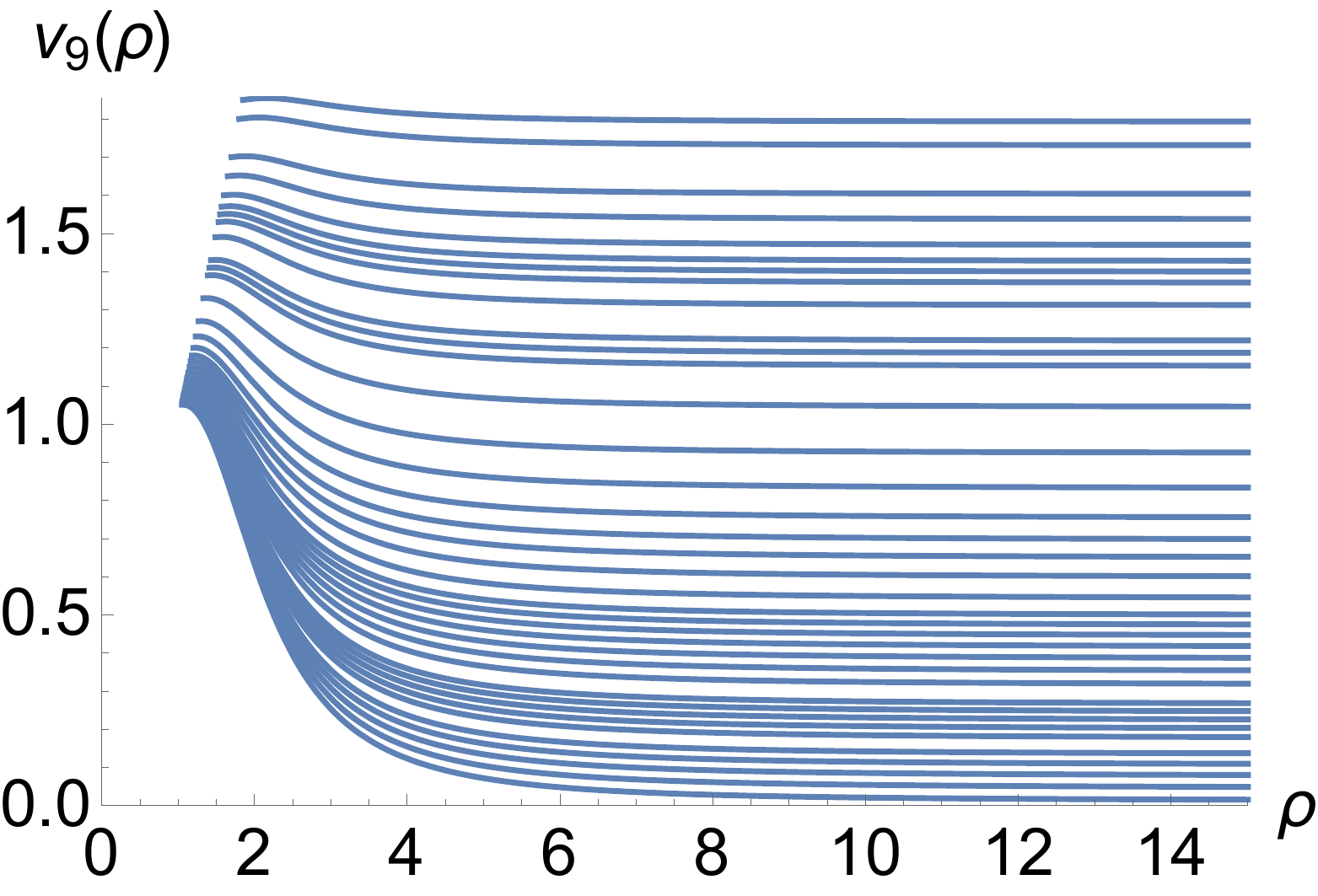} 
 \vspace{-0.4cm}
 
\noindent{{\textit{Figure 2: Numerical solutions for the vacuum functions $v_9(\rho)$.}}}
\end{center} \vspace{-0.3cm}

relevant for the vacuum configuration becomes
\beq S_{D7} \sim -  \int d^4x ~d\rho ~ {1 \over {\cal C} K \rho}  \sqrt{1  + {  {\cal C}^2 K^5U_0^5\over 32 } \rho^5 (\partial_\rho v_9)^2  }  \eeq
 The large $\rho$ solution is $v_9 = m + c/\rho^3$. Note here we use little $m$ and $c$ - they are masses and condensates between the chiral fermions on the domain wall which are distinct from the $M,C$ of the 4+1d theory. The condensate is identified by taking a derivative with respect to the mass, $m U_0$, on the action - as written in this limit the action is $m$ independent so one must imagine a sub-leading term, for example coming from the expansion of the dilaton, $\int d \rho ~v_9^2 / {\cal C} K \rho$. Now one sets $v_9 = m + c/\rho^3$ and differentiates the cross term w.r.t $U_0 m$:  thus we find the condensate is proportional to  $c~ / {\cal C}  K U_0$ which is both proportional to $c$ and of dimension 3. 

The resulting full equation of motion for an $x$ independent $v_9$ vacuum solution is
\beq \begin{array}{l}\partial_\rho \bigg(e^{-\phi}\rho^2 {G_x^{5/2}G_v^{3/2}\over G_v^{1/2}(\rho) }(\partial_\rho x_4) {{\cal F}\; \partial_\rho v_9\over \sqrt{1+{\cal F}(\partial_\rho v_9)^2}}\bigg) \\ \\ -{2\rho^2 (\partial_\rho x_4)\over G_v^{1/2}(\rho)}v_9 {\partial\over\partial v^2}\Big(e^{-\phi} G_x^{5/2}G_v^{3/2}\sqrt{1+{\cal F}(\partial_\rho v_9)^2}\Big)=0\end{array} \label{vacv9}\eeq

In the UV the solution is of the form $m + c/\rho^2$. We find solutions numerically by shooting from the IR boundary conditions $v_9(\rho_{\rm min}) = \rho_{\rm min}$ (this is required for the IR mass gap to be consistent with the gap described by the loci in Figure 1) and $v_9'(\rho_{\rm min}) = 0$. We display the results in Figure 2. The numerics become highly tuned as $\rho_{\rm min}$ approaches one and the U-shaped loci become infinitely wide but the results look very consistent with the UV quark mass being zero in this limit (which is the case for the D7 embedding in a uncompactified D5 background). For small separations of the domain walls, large $\rho_{\rm min}$, the quark mass scales as $1/\rho_{\rm min}$ as we found in similar configurations in \cite{CruzRojas:2021pql}. The massless embedding shows chiral symmetry breaking behaviour generating the $\rho_{\rm min}=1$ mass gap.

\begin{center}
    \includegraphics[width=8cm]{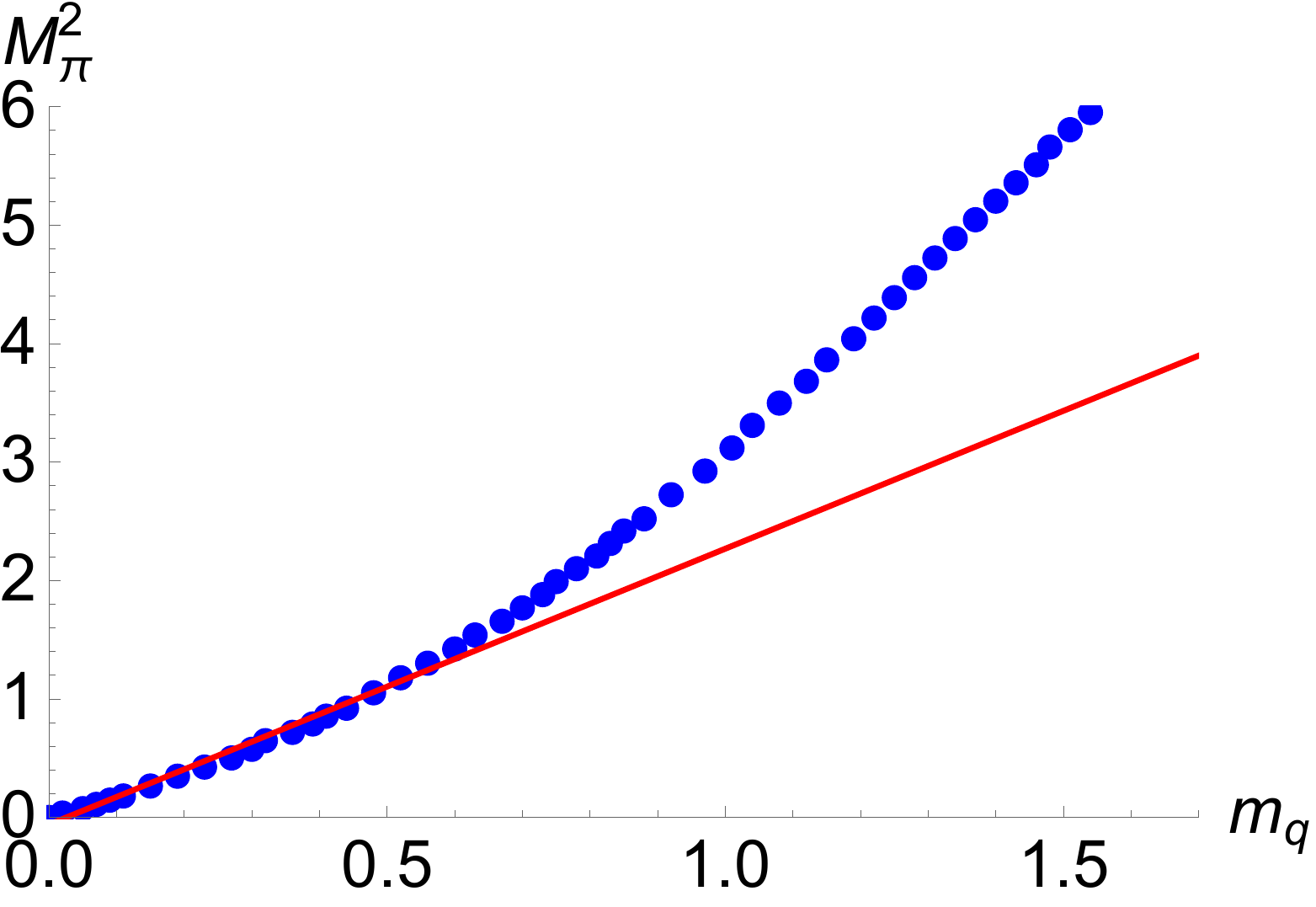} 
 \vspace{-0.4cm}
 
\noindent{{\textit{Figure 3: A plot of $M_\pi^2$ against $m_q$ with a guiding linear function plotted (red).}}}
\end{center} \vspace{-0.5cm}

\subsection{Pions} \vspace{-0.6cm}

The quark condensate and mass are complex objects and we would expect a second degree of freedom in the dual that forms a complex pair with $v_9$. Let us call this $v_{10}$ although there is no such field in the DBI action. We can immediately write down it's equation following that for $v_9$ since it has a U(1) symmetry that mixes it with that 
field. The equation of motion for fluctuations of $v_{10}$ in the background of the $v_9$ vacuum solution is simply
\beq \begin{array}{l}\partial_\rho \bigg(e^{-\phi}\rho^2 {G_x^{5/2}G_v^{3/2}\over G_v^{1/2}(\rho) }(\partial_\rho x_4) {{\cal F}\; \partial_\rho v_{10}\over \sqrt{1+{\cal F}(\partial_\rho v_{9})^2}}\bigg) \\ \\ -{2\rho^2 (\partial_\rho x_4)\over G_v^{1/2}(\rho)}v_{10} {\partial\over\partial v^2}\Big(e^{-\phi} G_x^{5/2}G_v^{3/2}\sqrt{1+{\cal F}(\partial_\rho v_9)^2}\Big)\\ \\
+ M^2 e^{-\phi}\rho^2 {G_x^{3/2}G_v^{5/2}\over G_v^{1/2}(\rho) }(\partial_\rho x_4) { v_{10}\over \sqrt{1+{\cal F}(\partial_\rho v_{9})^2}} =0\end{array} \label{pionem}\eeq
This equation is therefore sufficient to compute the behaviour of the Goldstone mode and its radially excited states of the theory. $v_{10}$ does not appear explicitly in the 
model but this is because the $v_9+ i v_{10}$ complex number can be written as $v_9 e^{i \phi}$ and then a U(1)$_A$ transformation used to set $\phi=0$. The degrees of freedom though remain and the solutions will emerge as components of the gauge fields which are present on the U-shaped locus. It is easiest to compute using the logic here though. 

The Goldstone nature of this $v_{10}$ state follows simply from  (\ref{pionem}). If one sets $M^2=0$ and $v_{10}$ equal to the $v_9$ background solution then  (\ref{pionem}) is simply (\ref{vacv9}). This solution though can only be used as a physical state for the massless theory since we require that asymptotically it falls to zero so it describes a fluctuation of the operator (rather than asymptoting to a source). Away from the massless quark theory we must solve (\ref{pionem}) numerically with $v'_{10}(\rho_{\rm min})=0$ and vary $M^2$ to achieve $v_{10}(\infty)=0$. We show our numerical data in Figure 3. The results sensibly match a Gell-Mann-Oakes-Renner relation ($M_\pi^2 \sim m_q$)\cite{GellMann:1968rz} at small quark mass but at larger quark mass $M^2 \sim m_q^2$ as one would expect.

\subsection{Vector and Axial vector mesons}

The Lagrangian for a small $F^{\mu\nu}$ fluctuation is given by
\beq {\cal L} = {{\cal N} \over 2} \rho^2  {e^{-\phi} G_x^{5/2} G_v^{3/2} \over G_v^{1/2}(\rho)}{(\partial_\rho x_4)\over \sqrt{1  + {\cal F}(\partial_\rho v_9)^2 } }g^{\mu \alpha} g^{\nu \beta} F_{\mu \nu}F_{\alpha \beta} 
\eeq
Note here the $x_4$ derivative should be included in the sense that on the vacuum locus it contributes to the $\rho$ derivative as $\partial_{x^4} = \partial_{\rho}/ \partial_{\rho} x^4$.
The resulting equation for the spatial mesons is given by
\beq \label{va} \begin{array}{r}\partial_\rho \left[ \rho^2 e^{-\phi} G_x^{3/2} {G_v^{1/2} \over G_v^{1/2}(\rho)}{(\partial_\rho x_4) {\cal F}\over \sqrt{1  + {\cal F}(\partial_\rho v_9)^2 } } \partial_\rho A_x\right] \\ \\+ M^2 \rho^2 e^{-\phi} G_x^{1/2} {G_v^{3/2} \over G_v^{1/2}(\rho)}{(\partial_\rho x_4)\over \sqrt{1  + {\cal F}(\partial_\rho v_9)^2 } } A_x = 0 
\end{array}\eeq
Vector mesons have IR boundary conditions $\partial_\rho A(\rho_{\rm min})=0$ and $A(\infty)=0$ (note the linearized equation doesn't depend on $A(\rho_{\rm min})$ so one varies $M^2$ to satisfy the UV boundary condition). Axial-vector mesons have $A(\rho_{\rm min})=0$ and $A(\infty)=0$ (again the linearity means the derivative is only defined up to a constant - so one picks some fixed IR derivative and varies $M^2$ to find a solution that matches the UV boundary conditions). 

The solutions for the vector meson must be normalized - one requires that it's kinetic term is canonical so 
\beq {\cal N}  \int d \rho~ \rho^2  {e^{-\phi} G_x^{1/2} G_v^{3/2} \over G_v^{1/2}(\rho)}{(\partial_\rho x_4)\over \sqrt{1  + {\cal F}(\partial_\rho v_9)^2 } }  (A_x)^2 = 1  \eeq

To normalize the source solutions we must investigate the UV behaviour of (\ref{va}). At large $\rho$ we have
\beq
\partial_\rho \left[ {C K^4 U_0^3 \over 8}  \rho^4 \partial_\rho A_x\right] + M^2 {4 \over {\cal C} K^3 U_0^2 } {1 \over \rho^3}A_x = 0 
\eeq
The solutions of this are not of the Log $Q^2/\rho^2$ form found in AdS/QCD \cite{Erlich:2005qh} since the UV of the theory is not a conformal 3+1d theory (the higher dimensional glue theory's coupling runs as a power law in the UV). However, it is always a sleight of hand to match a gravity dual to perturbative QCD since the dual must fail (or become strongly coupled itself) as QCD becomes perturbative. A simple fix is to only allow the gravity description to extend to a UV cut off. We will take $10 U_0$ - $U_0$ sets the scale of the IR quark mass as shown in Figure 2 so is matched to of order 300 MeV - thus the UV cut off scale corresponds 3 GeV or so. One should match to QCD at this UV cut off in the region where QCD is between weak and strong coupling. Rather than attempting to match (which would require calculation in  QCD in an intermediate coupling regime) we will simple set the normalization of the source solutions by fitting to $F_V$. We can then predict $F_A$ and $f_\pi$. 

$F_V^2$ is defined as the Feynman rule for a vector meson to turn directly into it's source at $q^2=0$. We must solve (\ref{va}) with $M^2=0$ to find a solution that asymptotes to a constant in the UV to represent the source. 

Now we can compute the decay constant (removing the UV surface term between the operator and source) as
\beq \begin{array}{c}
F_V^2= {\cal N} \int d\rho~ \partial_\rho\left[\rho^2  {e^{-\phi} G_x^{3/2} G_v^{1/2} \over G_v^{1/2}(\rho)}  \right. \hspace{4cm}\\ \\ \left. ~~~~~~~~ \times {(\partial_\rho x_4) {\cal F}\over \sqrt{1  + {\cal F}(\partial_\rho v_9)^2 } } \partial_\rho A_V \right]  A_{\rm source}   \end{array}
\eeq
We cut the integration off at $10 U_0$ and set the source normalization to give the observed value of $F_V$. The $F_A$ coupling is then a repeat of this computation with the axial vector meson solutions and using the same normalization at the cut off. 

$f_\pi^2$ is given by the axial axial correlator
\beq \begin{array}{l} f_\pi^2 = {\cal N} \int d \rho~ \rho^2  {e^{-\phi} G_x^{3/2} G_v^{1/2} \over G_v^{1/2}(\rho)} \\ \\ \left. \right. \hspace{2cm} \times
{(\partial_\rho x_4) {\cal F}\over \sqrt{1  + {\cal F}(\partial_\rho v_9)^2 } }(\partial_\rho A_{\rm source ~A})^2 \end{array}\eeq

Now we can compute $m_\pi$, $M_V$, $M_A$, $F_V$, $F_A$ and $f_\pi$. We use $M_V$ and $M_\pi$ to set the overall scale and quark mass and $F_V$ sets the source normalization, leaving 3 predictions. We display these results in Table 1.  The model like many AdS/QCD models gives the correct ball-park spectrum but here we find the axial sector predictions ($f_\pi, M_A, F_A$) all lie above the QCD values. Radially excited states' masses also rise sharply suggesting $M \sim n$ rather than $M \sim \sqrt{n}$ as is widely the case in AdS/QCD models \cite{Shifman:2005zn}.

We can improve the predictions by adding higher dimension operators at the UV cut off scale \cite{Evans:2006ea}. These should represent the generation of such operators in the intermediate regime between strong and weak coupling where one should match to perturbative QCD. Using Witten's multi-trace prescription, we change the UV boundary conditions on the holographic fields to allow solutions with non-zero source. We interpret the source as due to the presence of generically an operator $G {\cal O}^\dagger {\cal O}$ which when ${\cal O}$ condenses  generates an effective source $G \langle {\cal O} \rangle$.

\begin{center}
\begin{tabular}{c|c|c|c}
& QCD & DW AdS/QCD & Improved \\
&&& DW AdS/QCD   \\ \hline
&&&\\
$m_\rho$ & 775 MeV & $775^*$& $g_q =0.247$\\
$m_\pi$ & 139 MeV & $139^*$ & $g_v=0.656$ \\
$m_a$ & 1230 MeV & $1,955$ & $g_A=1.287$\\
$F_V$ & 345 MeV & $345^*$& \\
$F_A$ & 433 MeV  &$726.7$ &\\
$f_\pi$ &93 MeV &$135.3$ & $128.8$\\
&&&\\
$M_{v,n=1}$&1465 MeV& 3284 &1881.8  \\ $M_{A,n=1}$ & 1655 MeV &5043&2752.5 \\&&&\\\hline
\end{tabular}

\noindent{{\textit{Table 1: Mesonic observables - QCD values and the basic Domain Wall AdS/QCD model's predictions. Starred quantities are used to fix parameters as described in the text. In the final column we list the values of the higher dimension operator couplings in the improved version of the model - here $f_\pi$, and the excited state masses are predicted.}}}
\end{center}

 See \cite{Clemens:2017udk} for recent examples of this methodology in alternative AdS/QCD set ups.  

In particular we proceed as follows.  We start by considering different background embeddings for $v_9$ that asymptote in the UV to different source values. For each we compute the pion mass. We then fix by hand the ratio of the vector meson mass to the pion mass to its observed value and find the wave function, which does not asymptote to zero in the UV - we can extract the HDO coupling from the source and operator values at the cut off, assuming the presence of an operator $g_V^2/ \Lambda^2 |\bar{q} \gamma^\mu q|^2$ (we will quote $g_V^2= \Lambda^2 {\cal J}/{\cal O}$). Next we fit the normalization of the source to fit $F_V$. In the axial sector we allow a coupling $g_A^2/ \Lambda^2 |\bar{q} \gamma^\mu \gamma^5 q|^2$ to fit the axial vector meson mass. Now $F_A$ and $f_\pi$ can be computed. Repeating this for all the $v_9$ embeddings we can achieve the physical value of $f_A$, fixing the background embedding. The pion decay constant reduces a little as shown in Table 1 but not as low as the physical value. There is a bigger improvement in the predictions of the radial excited state masses as we show for the first excitations of the $\rho$ and $a$ mesons, although they too still remain high.

\section{III Discussion}

We have presented a holographic domain wall theory of 3+1 dimensional  chiral quarks interacting via confining gauge interactions. Here the gauge interactions are five dimensional albeit with one compact dimension to generate the confinement scale. The quarks of a 4+1 dimensional theory are isolated on separated domain walls where the 4+1 dimensional theory's mass vanishes. The holographic fields on the locus of the defects provide a holographic description of a QCD-like theory. We have shown the theory has chiral symmetry breaking and generates a spectrum that quite closely resembles QCD. Deviations are likely due to the gauge coupling growing into the UV - we have included a UV cut off to stop this growth and  included some higher dimension operators at the cut off. The spectrum is then improved but the full effects of the higher dimension gauge dynamics are not suppressed. 

In lattice simulations using the domain wall fermion method the gauge fields are isolated on the defects and independent of the higher dimensions. It would be interesting to try to arrange such a set up holographically using multi-centre brane solutions, although non-supersymmetric multi-centre solutions are hard to find. 

We have presented the model on the surface of a single D7 brane generating just a single flavour of quarks. However, one would expect the domain wall trick to generate non-abelian  SU($N_f)_L \times$ SU($N_f)_R$ flavour symmetries - on a domain wall only a single chiral quark is massless whilst the other is massive, so the interaction with the adjoint scalar superpartner of the gauge field is suppressed on the wall. Thus the theory on the surface of $N_f$ D7 branes is just that of the abelian case but fields are promoted to $N_f \times N_f$ matrices and the full action should be traced in flavour space. The bosonic fields  will form U($N_f$) multiplets of the vector flavour symmetry with the masses and couplings of the abelian case we have described. 

In conclusion we believe it has been interesting to generate a new type of AdS/QCD model which uses the domain wall fermion method. The method may allow a wider class of chiral theories to be explored in the future.

\noindent {\bf Acknowledgements:} 
NEs work was supported by the STFC consolidated grants ST/P000711/1 and ST/T000775/1. JMs work was supported by an STFC studentship.



\begin{thebibliography}{ll}
\bibitem{Kaplan:1992bt}
D.~B.~Kaplan,
Phys. Lett. B \textbf{288} (1992), 342-347
doi:10.1016/0370-2693(92)91112-M
[arXiv:hep-lat/9206013 [hep-lat]].

\bibitem{CruzRojas:2021pql}
J.~Cruz Rojas, N.~Evans and J.~Mitchell,
[arXiv:2106.08753 [hep-th]].

\bibitem{Maldacena:1997re}
  J.~M.~Maldacena,
  Adv.\ Theor.\ Math.\ Phys.\  {\bf 2} (1998) 231  [hep-th/9711200]; E.~Witten,
  Adv.\ Theor.\ Math.\ Phys.\  {\bf 2} (1998) 253  [hep-th/9802150];  S.~S.~Gubser, I.~R.~Klebanov and A.~M.~Polyakov,
  Phys.\ Lett.\ B {\bf 428} (1998) 105  [hep-th/9802109].  
  
  
\bibitem{Erlich:2005qh}
J.~Erlich, E.~Katz, D.~T.~Son and M.~A.~Stephanov,
Phys. Rev. Lett. \textbf{95} (2005), 261602
doi:10.1103/PhysRevLett.95.261602
[arXiv:hep-ph/0501128 [hep-ph]]; L.~Da Rold and A.~Pomarol,
Nucl. Phys. B \textbf{721} (2005), 79-97
doi:10.1016/j.nuclphysb.2005.05.009
[arXiv:hep-ph/0501218 [hep-ph]].

\bibitem{Sakai:2004cn}
T.~Sakai and S.~Sugimoto,
Prog. Theor. Phys. \textbf{113} (2005), 843-882
doi:10.1143/PTP.113.843
[arXiv:hep-th/0412141 [hep-th]].



\bibitem{Evans:2006ea}
N.~Evans and A.~Tedder,
Phys. Lett. B \textbf{642} (2006), 546-550
doi:10.1016/j.physletb.2006.10.019
[arXiv:hep-ph/0609112 [hep-ph]].; N.~Evans, J.~P.~Shock and T.~Waterson,
Phys. Lett. B \textbf{622} (2005), 165-171
doi:10.1016/j.physletb.2005.07.014
[arXiv:hep-th/0505250 [hep-th]].



\bibitem{Witten:2001ua}
E.~Witten,
[arXiv:hep-th/0112258 [hep-th]].

\bibitem{Evans:2016yas}
N.~Evans and K.~Y.~Kim,
Phys. Rev. D \textbf{93} (2016) no.6, 066002
doi:10.1103/PhysRevD.93.066002
[arXiv:1601.02824 [hep-th]].

\bibitem{Clemens:2017udk}
W.~Clemens and N.~Evans,
Phys. Lett. B \textbf{771} (2017), 1-4
doi:10.1016/j.physletb.2017.05.027
[arXiv:1702.08693 [hep-th]];
J.~Erdmenger, N.~Evans, W.~Porod and K.~S.~Rigatos,
JHEP \textbf{02} (2021), 058
doi:10.1007/JHEP02(2021)058
[arXiv:2010.10279 [hep-ph]];
M.~Jarvinen,
JHEP \textbf{07} (2015), 033
doi:10.1007/JHEP07(2015)033
[arXiv:1501.07272 [hep-ph]].

\bibitem{Myers:2006qr}
R.~C.~Myers and R.~M.~Thomson,
JHEP \textbf{09} (2006), 066
doi:10.1088/1126-6708/2006/09/066
[arXiv:hep-th/0605017 [hep-th]].

\bibitem{Itzhaki:1998dd}
N.~Itzhaki, J.~M.~Maldacena, J.~Sonnenschein and S.~Yankielowicz,
Phys. Rev. D \textbf{58} (1998), 046004
doi:10.1103/PhysRevD.58.046004
[arXiv:hep-th/9802042 [hep-th]].

\bibitem{Horowitz:1998ha}
G.~T.~Horowitz and R.~C.~Myers,
Phys. Rev. D \textbf{59} (1998), 026005
doi:10.1103/PhysRevD.59.026005
[arXiv:hep-th/9808079 [hep-th]].

\bibitem{Babington:2003vm}
J.~Babington, J.~Erdmenger, N.~J.~Evans, Z.~Guralnik and I.~Kirsch,
Phys. Rev. D \textbf{69} (2004), 066007
doi:10.1103/PhysRevD.69.066007
[arXiv:hep-th/0306018 [hep-th]].





  
  \bibitem{Karch:2002sh}
  A.~Karch and E.~Katz,
  JHEP {\bf 0206} (2002) 043
  [arXiv:hep-th/0205236]; M.~Grana and J.~Polchinski,
  Phys.\ Rev.\  D {\bf 65} (2002) 126005
  [arXiv:hep-th/0106014]; M.~Bertolini, P.~Di Vecchia, M.~Frau, A.~Lerda and R.~Marotta,
  Nucl.\ Phys.\  B {\bf 621} (2002) 157
  [arXiv:hep-th/0107057];M.~Kruczenski, D.~Mateos, R.~C.~Myers and D.~J.~Winters,
  JHEP \textbf{07}, 049 (2003)
  doi:10.1088/1126-6708/2003/07/049
  [arXiv:hep-th/0304032 [hep-th]];
  J.~Erdmenger, N.~Evans, I.~Kirsch, and E.~Threlfall,   {\em Eur. Phys. J.} {\bf A35} (2008)
  81--133, [arXiv:0711.4467].
  
 



\bibitem{GellMann:1968rz}
M.~Gell-Mann, R.~J.~Oakes and B.~Renner,
Phys. Rev. \textbf{175} (1968), 2195-2199
doi:10.1103/PhysRev.175.2195

\bibitem{Shifman:2005zn}
M.~Shifman,
doi:10.1142/9789812774132\_0025
[arXiv:hep-ph/0507246 [hep-ph]]; A.~Karch, E.~Katz, D.~T.~Son and M.~A.~Stephanov,
Phys. Rev. D \textbf{74} (2006), 015005
doi:10.1103/PhysRevD.74.015005
[arXiv:hep-ph/0602229 [hep-ph]].


\end{thebibliography}
\end{document}